\documentclass[twocolumn,showpacs,preprintnumbers,superscriptaddress,amsmath,amssymb]{revtex4}
\usepackage{amssymb}		
\usepackage{amsmath}
\usepackage{graphicx}
\usepackage[normalem]{ulem}
\usepackage{multirow}
\usepackage{appendix}
\usepackage{CJK}
\usepackage[usenames]{color}
\usepackage{bm}
\usepackage{hyperref}
\usepackage{booktabs} 	
\usepackage{leftidx}

\begin{document}
\begin{CJK*} {UTF8} {gbsn}

\title{Nuclear system size scan for freeze-out properties in relativistic heavy-ion collisions by using a multiphase transport model}

\author{Dong-Fang Wang}

\affiliation{Shanghai Institute of Applied Physics, Chinese Academy of Sciences, Shanghai 201800, China}
\affiliation{Key Laboratory of Nuclear Physics and Ion-beam Application (MOE), Institute of Modern Physics, Fudan University, Shanghai 200433, China}
\affiliation{University of Chinese Academy of Sciences, Beijing 100049, China}

\author{Song Zhang}\thanks{Email: song\_zhang@fudan.edu.cn}
\affiliation{Key Laboratory of Nuclear Physics and Ion-beam Application (MOE), Institute of Modern Physics, Fudan University, Shanghai 200433, China}

\author{Yu-Gang Ma}\thanks{Email:  mayugang@fudan.edu.cn}
\affiliation{Key Laboratory of Nuclear Physics and Ion-beam Application (MOE), Institute of Modern Physics, Fudan University, Shanghai 200433, China}
\affiliation{Shanghai Institute of Applied Physics, Chinese Academy of Sciences, Shanghai 201800, China}

\begin{abstract}
	 A system size scan program was recently proposed for the STAR experiments at the Relativistic Heavy Ion Collider(RHIC). In this study, we employ a multiphase transport (AMPT) model for considering the bulk properties at the freeze-out stage for  $\mathrm{^{10}B+^{10}B}$, $\mathrm{^{12}C+^{12}C}$, $\mathrm{^{16}O+^{16}O}$, $\mathrm{^{20}Ne+^{20}Ne}$, $\mathrm{^{40}Ca+^{40}Ca}$, $\mathrm{^{96}Zr+^{96}Zr}$, and $\mathrm{^{197}Au+^{197}Au}$ collisions at RHIC energies $\sqrt{s_{NN}}$ of 200, 20, and 7.7 GeV. The results for $\mathrm{^{197}Au+^{197}Au}$ collisions are comparable with those of previous experimental STAR data. The transverse momentum $p_{T}$ spectra of charged particles ($\pi^{\pm}$, $K^{\pm}$, $p$, and $\bar{p}$) at the kinetic freeze-out stage, based on a blast-wave model, are also discussed. In addition, we use a statistical thermal model to extract the parameters at the chemical freeze-out stage, which agree with those from other thermal model calculations. It was found that there is a competitive relationship between the kinetic freeze-out parameter $T_{kin}$ and the radial expansion velocity $\beta_{T}$, which also agrees with the STAR or ALICE results. We found that the chemical freeze-out strangeness potential $\mu_{s}$ remains constant in all collision systems and that the fireball radius $R$ is dominated by $\left\langle \mathrm{N_{Part}}\right\rangle$, which can be well fitted by a function of $a \left\langle \mathrm{N_{Part}}\right\rangle^{b}$ with $b \approx 1/3$. In addition, we calculated the nuclear modification factors for different collision systems with respect to the $ \mathrm{{}^{10}B} + \mathrm{{}^{10}B}$ system, and found that they present a gradual suppression within a higher $p_{T}$ range from small to large systems.
 	\end{abstract}
\maketitle

	\section{Introduction}
	\par
	Over the past few years, there have been numerous efforts to explore a quantum chromodynamics (QCD) phase diagram and quark gluon plasma, which are important goals for ultra-relativistic heavy-ion collision experiments \cite{phase_e1,phase_e2,phase_e3,phase_e4,PBMPhysRep,ChenPhysRep,LuoNST,SongNST,MaYG}. A QCD phase diagram is characterized by temperature $(T)$ and the baryon chemical potential $(\mu_{B})$ \cite{PBMPhysRep,PBMNature}. Lattice QCD calculations predict a phase transition from a state of hadronic constituents, where the degrees of freedom are hadronic, to a plasma of deconfined quarks and gluons dominated by partonic degrees of freedom at a critical temperature of $T_{c} \approx 170$ MeV \cite{tc_1,tc_2}, namely quark-gluon plasma (QGP) \cite{QCD2}. QGP was found inside a hot and dense fireball created at the early stage of central Au + Au collisions at $\sqrt{s_{NN}} = 200$ GeV in the Relativistic Heavy-Ion Collider (RHIC) \cite{09STAR_bulk} at Brookhaven National Laboratory, as well as during Pb + Pb collisions at $\sqrt{s_{NN}} = 2.76$ TeV performed at the Large Hadron Collider (LHC), as reported through the ALICE Collaboration~\cite{13ALICE_276}. Many QGP signatures have been proposed based on simultaneous observations of different bulk quantities, which include the chemical freeze-out temperature $(T_{ch})$, baryon chemical potential $(\mu_{B})$, and kinetic freeze-out temperature $(T_{kin})$, as well as the average radial expansion velocity $(\beta_{T})$, which can be studied through the transverse momentum $(p_{T})$ spectra of the particles.
		
	The freeze-out properties provide evolution information on the collision system, which helps us to understand the expansion of the fireball \cite{PBMNature,Tfree,Tfree2}. The thermal model successfully describes the production of particles in heavy-ion collisions with a few parameters such as the chemical freeze-out temperature, baryon chemical potential, and fireball volume. From particle yields or ratios, the thermal model can be used to obtain the chemical freeze-out properties, such as the chemical freeze-out temperature $(T_{ch})$, as well as the baryon $(\mu_{B})$ and strangeness $(\mu_{S})$ chemical potentials \cite{sps_thermal}. Apart from the transport or thermal models, the blast-wave model developed through hydrodynamics has also been extremely successful in describing observables, such as identified particle transverse momentum $p_{T}$ spectra, up to a few GeV/c \cite{13ALICE_276}. By fitting the transverse momentum distribution, the blast-wave model has often been applied to extract the kinetic freeze-out properties, such as the kinetic freeze-out temperature and the radial flow velocity.
		
       	A system size scan program was recently proposed at RHIC energies. The system provides the chance to further verify the validity of relativistic hydrodynamics in different collision systems \cite{otherssds}.
       	In this study, scans of $AA$ collision systems in the most central collisions occurring at the center of mass with energies of $\sqrt{s_{NN}}$ = 200, 20 and 7.7 GeV,
       namely, $\mathrm{\leftidx{^{10}}B} + \mathrm{\leftidx{^{10}}B}$, $\mathrm{\leftidx{^{12}}C} + \mathrm{\leftidx{^{12}}C}$,
       	$\mathrm{\leftidx{^{16}}O}+\mathrm{\leftidx{^{16}}O}$, $\mathrm{\leftidx{^{20}}Ne}+\mathrm{\leftidx{^{20}}Ne}$,
       	$\mathrm{\leftidx{^{40}}Ca}+\mathrm{\leftidx{^{40}}Ca}$, $\mathrm{\leftidx{^{96}}Zr}+\mathrm{\leftidx{^{96}}Zr}$,
       	$\mathrm{\leftidx{^{197}}Au}+\mathrm{\leftidx{^{197}}Au}$, were simulated using a multiphase transport (AMPT) model to provide some predictions of the parameters at freeze-out stage.
       	We present the AMPT prediction of $p_{T}$ and $dN/dy$ spectra of identified particles including $\pi^{\pm}$, $k^{\pm}$, $p$, and $\bar{p}$
       	in different symmetric collision systems. Furthermore, we investigate the system dependence of the freeze-out properties at the chemical and kinetic freeze-out stages \cite{sysscanrhic}.

	 The remainder of this paper is arranged as follows: In Sec.\ref{sec:model}, an introduction to a multiphase transport (AMPT) model and some input parameters
	used in this study are presented. The identified particle transverse momentum ($p_{T}$) spectra and yields $dN/dy$ are also given. The effects of different collision systems
	 on the freeze-out properties are also discussed in Sec.\ref{sec:freezeout}. Finally, a brief summary is presented in Sec.\ref{sec:summary}.

\section{Brief introduction to the AMPT model}
\label{sec:model}
		
		A multi phase transport  (AMPT) model~\cite{AMPT_origin}, which is a hybrid dynamic model, is employed to calculate different collision systems.
		The AMPT model can describe the $p_{T}$ distribution of charged particles \cite{xujun,suppressionhighpt,Ye_2017,JinXH,WangH} and their elliptic flow of Pb+Pb collisions
		at $\sqrt{s_{NN}} = 2.76$ TeV, as measured through the LHC-ALICE Collaboration. The model includes four main components to describe
		the relativistic heavy-ion collision process: the initial conditions simulated using the Heavy Ion Jet Interaction Generator (HIJING) model~\cite{HIJING-1,HIJING-2}, the partonic interactions described by Zhang's Parton Cascade (ZPC) model~\cite{ZPCModel}, the hadronization process through a Lund string fragmentation or coalescence model, and the hadronic re-scattering process using A Relativistic Transport (ART) model~\cite{ARTModel}. There are two versions of AMPT: 1) the AMPT version with a string melting mechanism, in which a partonic phase is generated from excited strings in the HIJING model, where a simple quark coalescence model is used to combine the partons into hadrons; and 2) the default AMPT version which only undergoes a pure hadron gas phase. The details of AMPT can be found in Ref.~\cite{AMPT_origin}.

		In the AMPT model, impact parameter $b$, which is the distance between the center of the two collided nuclei, can determine the collision centrality. In addition, the number of participants is always related to the centrality or impact parameter. In this study, we only focus on 0\%-5\% centrality events, the corresponding maximum impact parameters, the number of participants $N_{Part}$, and the number of events, which are listed in Table~\ref{AMPT_info}.
		
		In this calculation, we adopt the AMPT parameters, suggested in Ref.~\cite{Ye_2017}, and
		the select charged particles, $\pi^{\pm}$, $k^{\pm}$, $p$, and $\bar{p}$ with kinetic windows, $0.2<p_{T}<1.5$ GeV/c and $|y|<0.1$.
		
\section{Results and discussion}
\label{sec:freezeout}
	\subsection{Identified particle $p_{T}$ and $dN/dy$ spectra}
		
\begin{figure*}[htb]
				\includegraphics[angle=0,scale=0.9]{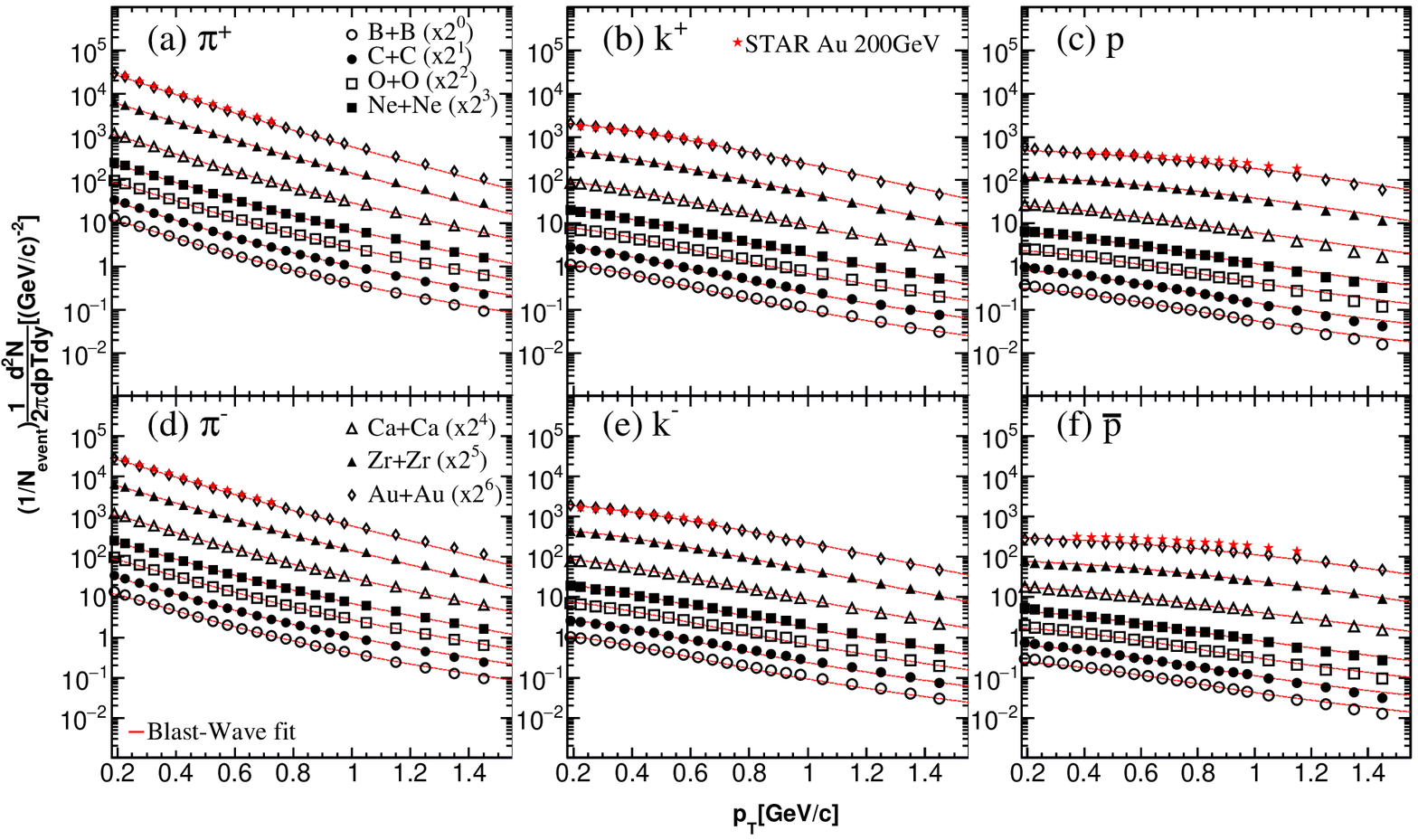}
				\caption{Transverse momentum $p_{T}$ spectra at midrapidity ($|y|<0.1$) for $\pi^{\pm}$, $k^{\pm}$, $p$, and $\bar{p}$ in $ \mathrm{{}^{10}B} + \mathrm{{}^{10}B}$, $ \mathrm{{}^{12}C} + \mathrm{{}^{12}C}$,
				$ \mathrm{{}^{16}O} + \mathrm{{}^{16}O}$, $ \mathrm{{}^{20}Ne} + \mathrm{{}^{20}Ne}$,
				$ \mathrm{{}^{40}Ca} + \mathrm{{}^{40}Ca}$, $ \mathrm{{}^{96}Zr} + \mathrm{{}^{96}Zr}$, and $ \mathrm{{}^{197}Au} + \mathrm{{}^{197}Au}$
				 collision systems at $\sqrt{s_{NN}} = 200$ GeV. The spectra are scaled by a factor of $2^{n}$, as marked in the inset
				(note that the data of the boron collision system are not scaled).
				The curves indicate that the blast-wave model combination fits the $0\%-5\%$
				centrality AMPT results for different collision systems. Experimental data are taken from the STAR Collaboration for $ \mathrm{Au} + \mathrm{Au}$ collisions at $\sqrt{s_{NN}} = 200$ GeV~\cite{09STAR_bulk}.}
				\label{BW_AMPT_com_new}
			\end{figure*}

Figure~\ref{BW_AMPT_com_new} shows the results of the transverse momentum spectra for $\pi^{\pm}$, $k^{\pm}$, $p$, and $\bar{p}$ in $\mathrm{\leftidx{^{10}}B} + \mathrm{\leftidx{^{10}}B}$, $\mathrm{\leftidx{^{12}}C} + \mathrm{\leftidx{^{12}}C}$, $\mathrm{\leftidx{^{16}}O}+\mathrm{\leftidx{^{16}}O}$, $\mathrm{\leftidx{^{20}}Ne}+\mathrm{\leftidx{^{20}}Ne}$, $\mathrm{\leftidx{^{40}}Ca}+\mathrm{\leftidx{^{40}}Ca}$, $\mathrm{\leftidx{^{96}}Zr}+\mathrm{\leftidx{^{96}}Zr}$, and $\mathrm{\leftidx{^{197}}Au}+\mathrm{\leftidx{^{197}}Au}$ collisions at $\sqrt{s_{NN}} = 200$ GeV using the AMPT model. It seems that our simulation results of an Au + Au collision at $\sqrt{s_{NN}}$ = 200 GeV can describe the experimental data reported by the STAR Collaboration for the transverse momentum spectra of $\pi^{\pm}$, $k^{\pm}$, $p$, and $\bar{p}$~\cite{09STAR_bulk}. The $p_T$ spectra in different collision systems always present an exponential-like distribution, and the slope for heavier particles (such as protons) is flatter (harder) than that for lighter particles (such as $\pi$'s) which is due to the so-called radial flow effect~\cite{PhysRevC.69.034909,1993_ssBW,09STAR_bulk}.

\begin{figure*}[htb]
				\includegraphics[angle=0,scale=0.9]{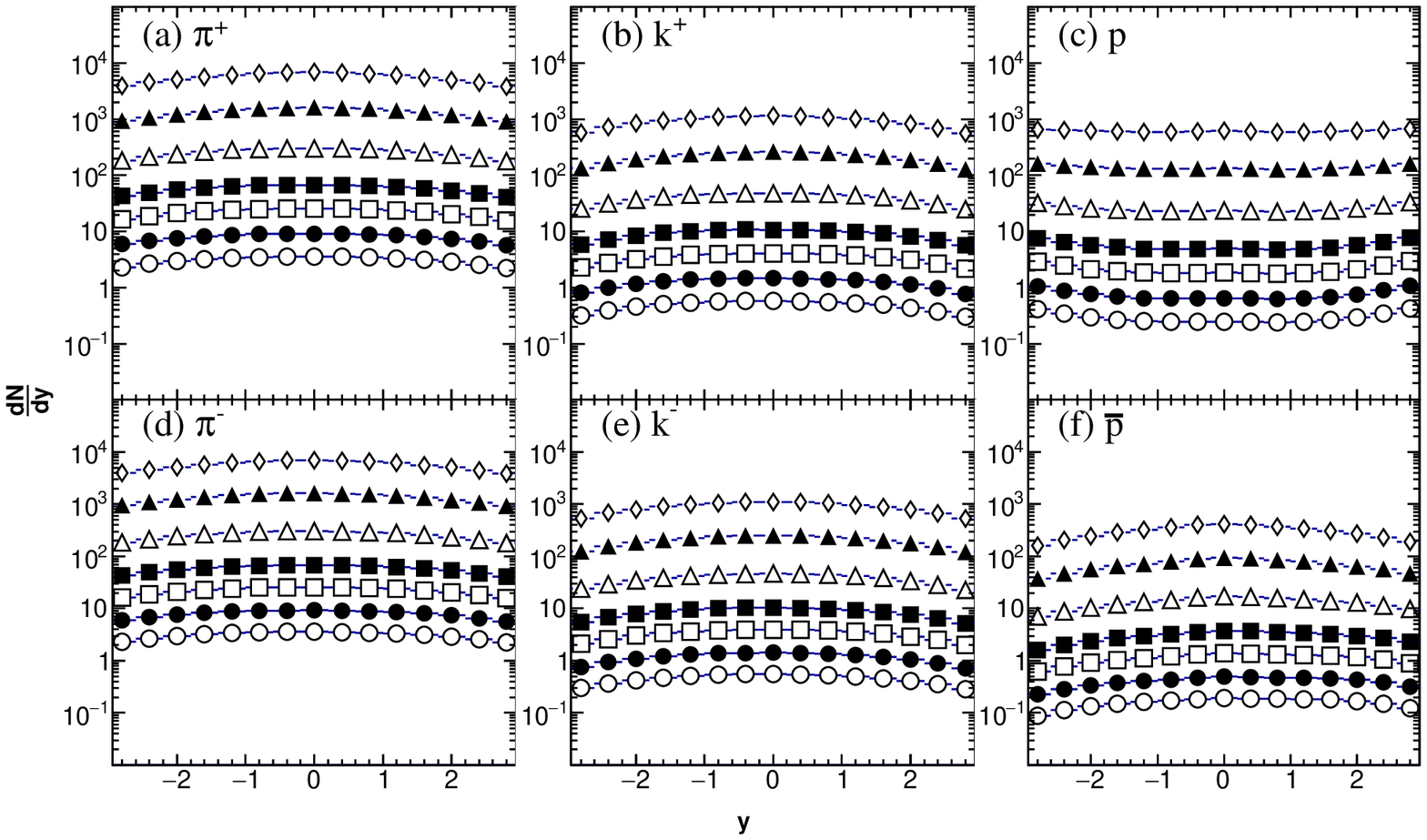}
				\caption{AMPT results of $dN/dy$ for identified $\pi^{\pm}$, $k^{\pm}$, $p$, and $\bar{p}$ yields in $ \mathrm{{}^{10}B} + \mathrm{{}^{10}B}$, $ \mathrm{{}^{12}C} + \mathrm{{}^{12}C}$,
				$ \mathrm{{}^{16}O} + \mathrm{{}^{16}O}$, $ \mathrm{{}^{20}Ne} + \mathrm{{}^{20}Ne}$, $ \mathrm{{}^{40}Ca} + \mathrm{{}^{40}Ca}$, $ \mathrm{{}^{96}Zr} + \mathrm{{}^{96}Zr}$, and
				$ \mathrm{{}^{197}Au} + \mathrm{{}^{197}Au}$ collision systems
				 at $\sqrt{s_{NN}} = 200$ GeV. }
				\label{dNdy_sys_dst}
			\end{figure*}

\begin{figure*}[htb]
				\includegraphics[angle=0,scale=0.9]{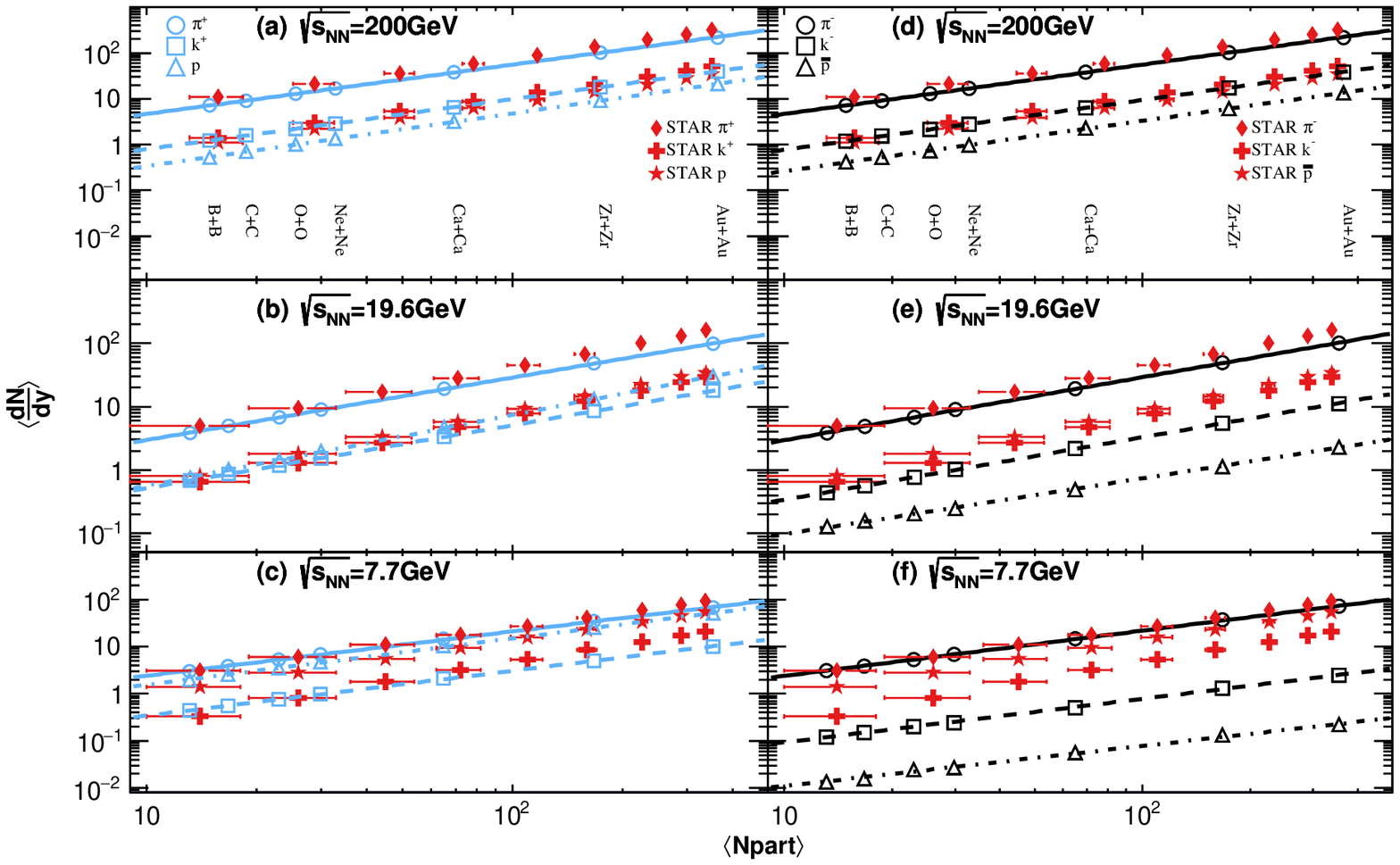}
				\caption{$\left\langle \mathrm{N_{part}}\right\rangle$ dependences of $\left\langle dN/dy\right\rangle$ for the identified $\pi^{\pm}$, $k^{\pm}$, $p$, and $\bar{p}$
				 at midrapidity $(|y|<0.1)$ in $ \mathrm{{}^{10}B} + \mathrm{{}^{10}B}$, $ \mathrm{{}^{12}C} + \mathrm{{}^{12}C}$, $ \mathrm{{}^{16}O} + \mathrm{{}^{16}O}$,
				 $ \mathrm{{}^{20}Ne} + \mathrm{{}^{20}Ne}$, $ \mathrm{{}^{40}Ca} + \mathrm{{}^{40}Ca}$, $ \mathrm{{}^{96}Zr} + \mathrm{{}^{96}Zr}$, and
				 $ \mathrm{{}^{197}Au} + \mathrm{{}^{197}Au}$ collision systems at $\sqrt{s_{NN}}$ = 200, 20, and 7.7 GeV.
				The solid line, dashed line, and dash-dotted line represent fits using a function of $\log_{10}(dN/dy) = p + q*\log_{10}(\left\langle \mathrm{N_{part}}\right\rangle)$
				for $\pi$, $k$, and $p$, respectively. The STAR Collaboration data for $ \mathrm{Au} + \mathrm{Au}$ collisions at $\sqrt{s_{NN}} = 200$ GeV are taken from Ref.~\cite{09STAR_bulk}.}
				\label{avg_dNdy_sys}
			\end{figure*}

	Figure~\ref{dNdy_sys_dst} shows $dN/dy$ of identified $\pi^{\pm}$, $k^{\pm}$, $p$, and $\bar{p}$ in the above mentioned collision systems at $\sqrt{s_{NN}}$ = 200 GeV, which keep a flat pattern at mid-rapidity. The $p_T$ and $dN/dy$ spectra present an obvious collision system dependence, i.e., the production yield increases with the size of the collision system. The average yield of particles $\left\langle {dN/dy} \right\rangle$ as a function of the average number of participants $\left\langle N_{part} \right\rangle$ is demonstrated in Fig.~\ref{avg_dNdy_sys} for $\pi^{\pm}$, $k^{\pm}$, $p$, and $\bar{p}$ at $\sqrt{s_{NN}}$ = 200, 20, and 7.7 GeV.
	The experimental data shown in Fig.~\ref{avg_dNdy_sys} are the  $\left\langle N_{part} \right\rangle$ dependences of the particle yields measured by the STAR Collaboration~\cite{09STAR_bulk} in Au+Au collisions at $\sqrt{s_{NN}}$ = 200 GeV. It can be seen that the results from the AMPT model are similar to those from the STAR experiments with a similar $\left\langle N_{part} \right\rangle$ although in different collision systems. At $\sqrt{s_{NN}}$ = 20 and 7.7 GeV, however, there are discrepancies between the AMPT simulation and STAR results for $k^{-}$ and $\bar{p}$, which leave room for a model improvement, particularly for the treatment of anti-particles. 

	\par
	The system ($\left\langle N_{part} \right\rangle$) dependence of $\left\langle dN/dy\right\rangle $ can be described through a simple function, $\log_{10}(dN/dy) = p + q*\log_{10}(\left\langle \mathrm{N_{part}}\right\rangle)$, and Table~\ref{dNdy_fit} shows the parameters of this fitting. The slope, demonstrated by $q$, is similar for particles and antiparticles, and larger for a heavier particle than for a lighter one.
	The experimental $dN/dy$ results can also be fitted by this type of function for $\pi$ and $k$ particles with extremely close values of $p$ and $q$. 
		 Parameter $q$ reflects the degree of $\left\langle dN/dy\right\rangle $ dependence on $\left\langle N_{part} \right\rangle$. For a given particle, $q$ increases with $\sqrt{s_{NN}}$.

	\begin{table*}[]
\scriptsize
\centering
\caption{AMPT input parameters and $\left\langle \mathrm{N_{part}}\right\rangle$ values of different collision systems.}
\label{AMPT_info}
\begin{tabular}{cc|cc|cc|cc}
\toprule
\multicolumn{2}{c}{} & \multicolumn{2}{c}{$\sqrt{s_{NN}}$ = 200GeV} & \multicolumn{2}{c}{$\sqrt{s_{NN}}$ = 20GeV} & \multicolumn{2}{c}{$\sqrt{s_{NN}}$ = 7.7GeV} \\
\cmidrule(r){3-4} \cmidrule(r){5-6} \cmidrule(r){7-8}
System & $\mathrm{\it{b_{max}}[\it{fm}]}$
&$\left\langle \mathrm{N_{part}}\right\rangle$ &Event counts
&$\left\langle \mathrm{N_{part}}\right\rangle$ &Event counts
&$\left\langle \mathrm{N_{part}}\right\rangle$ &Event counts      \\
\hline
$\mathrm{\leftidx{^{10}}B} + \mathrm{\leftidx{^{10}}B}$		&1.15619		&14.8  &60$\times 10^{4}$    	&13.2  &12$\times 10^{4}$  &13.1  &16$\times 10^{4}$\\
$\mathrm{\leftidx{^{12}}C} + \mathrm{\leftidx{^{12}}C}$		&1.22864		&18.7  &40$\times 10^{4}$    	&16.8  &6$\times 10^{4}$ 	&16.7  &10$\times 10^{4}$\\
$\mathrm{\leftidx{^{16}}O}+\mathrm{\leftidx{^{16}}O}$		&1.35229		&25.5   &20$\times 10^{4}$	&23.1  &4$\times 10^{4}$	&23.0  &10$\times 10^{4}$ \\
$\mathrm{\leftidx{^{20}}Ne}+\mathrm{\leftidx{^{20}}Ne}$		&1.45671		&32.8  &10$\times 10^{4}$	&30.0  &4$\times 10^{4}$	&29.8  &2$\times 10^{4}$\\
$\mathrm{\leftidx{^{40}}Ca}+\mathrm{\leftidx{^{40}}Ca}$		&1.83534		&69.3  &6$\times 10^{4}$	&65.0  &1$\times 10^{4}$	&64.9  &1$\times 10^{4}$ \\
$\mathrm{\leftidx{^{96}}Zr}+\mathrm{\leftidx{^{96}}Zr}$ 		&2.45727		&174.2 &2$\times 10^{4}$	&167.3  &2$\times 10^{4}$	&166.9 &3$\times 10^{4}$\\
$\mathrm{\leftidx{^{197}}Au}+\mathrm{\leftidx{^{197}}Au}$		&3.1226		&364.1 &3$\times 10^{4}$	&354  &3$\times 10^{4}$	&353.8  &3$\times 10^{4}$\\
\bottomrule
\end{tabular}
\end{table*}
			
			\begin{table*}
			\caption{ $\left\langle dN/dy\right\rangle$ fitting parameters for different charged particles using $\log_{10}(dN/dy) = p + q*\log_{10}(\left\langle \mathrm{N_{part}}\right\rangle)$ fits
			within the $p_{T}$ range $0.2-1.5$ GeV/c.}
			\label{dNdy_fit}
			\centering 
				\begin{tabular}{cccccccc}
					\toprule
					$\sqrt{s_{NN}}$ & Parameter & $\pi^{+}$ & $\pi^{-}$ & $k^{+}$ &$k^{-}$& $p$&$\bar{p}$\\
					\hline
					200GeV &p & $-0.413\pm0.003$ & $-0.413\pm0.003$ & $-1.19\pm0.003$ & $-1.21\pm0.003$ &$-1.63\pm0.004$ & $-1.68\pm0.005$\\
					&q & $1.08\pm0.002$ & $1.08\pm0.002$ &$1.09\pm0.002$ &$1.09\pm0.002$ &$1.15\pm0.003$&$1.10\pm0.003$\\
					\hline
					20GeV &p & $-0.513\pm0.004$ & $-0.526\pm0.00452$& $-1.28\pm0.007$ &$-0.526\pm0.005$ &$-1.37\pm0.007$ &$-1.89\pm0.013$ \\
					&q & $0.987\pm0.003$ &$0.995\pm0.003$ &$0.993\pm0.004$ &$0.984\pm0.004$ &$1.120\pm0.004$ &$0.880\pm0.007$\\
					\hline
					7.7GeV &p & $-0.564\pm0.004$ & $-0.584\pm0.004$ & $-1.42\pm0.006$ & $-1.96\pm0.011$ &$-0.799\pm0.004$ &$-2.81\pm0.03$\\
					&q & $0.945\pm0.003$ & $0.962\pm0.003$& $0.952\pm0.004$&$0.924\pm0.006$&$0.989\pm0.003$&$0.85\pm0.017$\\
					\hline
					\bottomrule
				\end{tabular}
				
			\end{table*}
			
	\subsection{Kinetic properties}
	
	The kinetic freeze-out properties can be extracted from the $p_{T}$ spectra, which characterize the information of the systems at the kinetic freeze-out stage (i.e., as the elastic interaction of the particles stops). During this stage, the temperature and radial expansion velocity are the key parameters used to describe the system. The kinetic freeze-out parameters are obtained by fitting the $p_{T}$ spectra with a hydrodynamics-motivated blast-wave model. The model makes a simple assumption that the particles are locally thermalized at the kinetic freeze-out temperature and are moving with a common transverse flow velocity field. Under the assumption of a radially boosted thermal source with a kinetic freeze-out temperature $T_{kin}$ and a transverse radial flow velocity $\beta_{T}$, the $p_{T}$ distribution of the particles is given as follows\cite{1993_ssBW}:
			\begin{equation}
				\begin{aligned}
					\frac { 1 } { p _ { T } } \frac { d N } { d p _ { T } }& \propto \int _ { 0 } ^ { R } r d r m _ {  { T } } I _ { 0 }
					 \left( \frac { p _ {  { T } } \sinh \rho } { T _ { \mathrm { kin } } } \right) K _ { 1 }
					 \left( \frac { m _ {  { T } } \cosh \rho } { T _ { \mathrm { kin } } } \right),
				\end{aligned}
				\label{eq:blatWM}
			\end{equation}
			where the velocity profile $\rho$ is described by
			\begin{equation}
				\begin{aligned}
					\rho = \tanh ^ { - 1 } \beta _ { \mathrm { T } } = \tanh ^ { - 1 } \left( \left( \frac { r } { R } \right) ^ { n } \beta _ { s } \right).
				\end{aligned}
			\end{equation}
Here, $m_{T} = \sqrt{p_{T}^{2} + m^{2}}$ is the transverse mass, $I_0$ and $K_1$ are the modified Bessel functions, $r$ is the radial distance in the transverse plane, $R$ is the radius of the fireball, $\beta_{T}$ is the transverse expansion velocity, and $\beta_{s}$ is the transverse expansion velocity at the surface. From these equations, the average transverse expansion velocity $\left\langle\beta_{T}\right\rangle = \frac{n}{n + 2} \beta_{s}$ can also be derived. The free parameters in the fits are the freeze-out temperature $T_{kin}$, the average transverse velocity $\left\langle\beta_{T} \right\rangle$, and the exponent of the velocity profile $n$.

\begin{figure}[htb]
				\includegraphics[angle=0,scale=0.44]{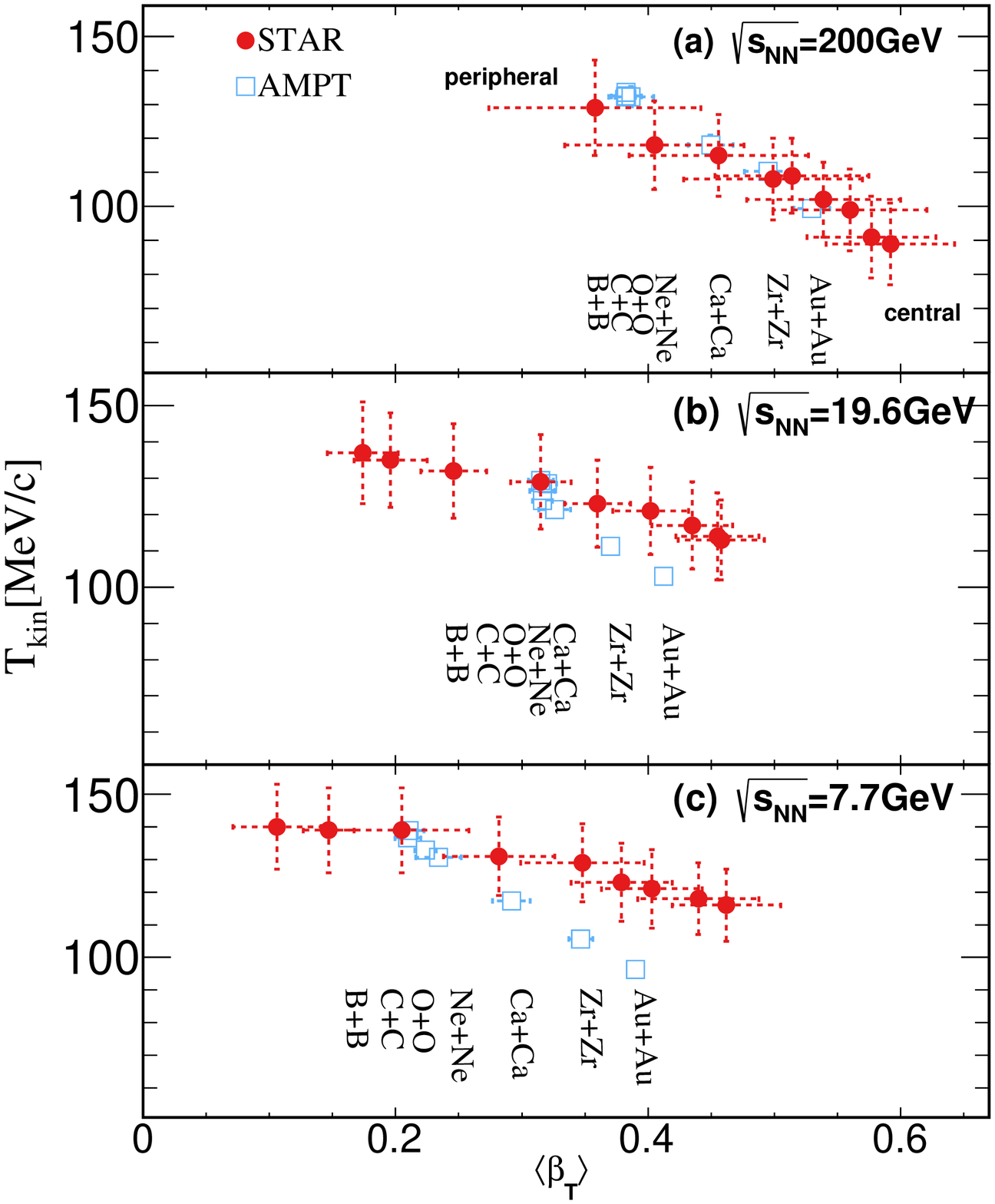}
				\caption{$T_{kin}$ as a function of $\left\langle\beta_{T}\right\rangle$ for different collision systems with the $\sqrt{s_{NN}}$ = 200, 20, and 7.7 GeV AMPT simulations shown in (a), (b), and (c), respectively. The STAR data for $ \mathrm{Au} + \mathrm{Au}$ collisions at $\sqrt{s_{NN}}$ = 200, 19.6, and 7.7 GeV are taken from Ref.~\cite{09STAR_bulk}}
				\label{Tot_flow_new}
			\end{figure}

Usually, $\pi^{\pm}$, $k^{\pm}$, $p$, and $\bar{p}$ particle spectra are fitted simultaneously with the blast-wave model rather than fitted individually \cite{13ALICE_276}. Figure~\ref{BW_AMPT_com_new} also presents the fitting using the blast-wave model with Eq.~(\ref{eq:blatWM}), and we can see that the $p_T$ spectra are fitted extremely well by this model.
			The extracted parameters, kinetic freeze-out temperature $T_{kin}$, and average radial flow velocity $\left\langle\beta_{T}\right\rangle$ are shown in Fig.~\ref{Tot_flow_new}.
			The system dependence of the fitted radial flow and the kinetic freeze-out temperature are both similar to the centrality ($\left\langle \mathrm{N_{part}}\right\rangle$) dependence of the parameters from the STAR Collaboration~\cite{09STAR_bulk} in Au+Au collisions at $\sqrt{s_{NN}}$ = 200 GeV, as shown in panel (a) of Fig.~\ref{Tot_flow_new}. The kinetic freeze-out temperature $T_{kin}$ decreases with the increase in the collision-system size, and the average radial flow velocity $\left\langle\beta_{T}\right\rangle$ presents an upward trend of the system dependence.
			However, at lower energies (20 and 7.7 GeV), as shown in Fig.~\ref{Tot_flow_new} (b) and (c), discrepancies are shown between the AMPT simulation and the STAR results~\cite{17STAR_bulk}.

	\subsection{Chemical properties}
	
	The system will reach the chemical freeze-out stage when inelastic collisions cease among the particles, that are created during the early stage. The chemical freeze-out properties provide information regarding the chemical equilibrium, such as the chemical freeze-out temperature $T_{ch}$, chemical freeze-out potential $\mu_B$ (baryon potential), and $\mu_s$ (strangeness potential), which determine the relative particle yield (particle ratio) in nucleus-nucleus collisions. Two approaches are typically used to obtain the chemical freeze-out parameters: a grand-canonical ensemble $(\mathrm{GCE})$ and a strangeness canonical ensemble $(\mathrm{SCE})$~\cite{Tch_fun}. In this study, we only consider the grand canonical case. For a hadron gas with volume $V$ and temperature $T$, the logarithm of the total partition function is given by the following \cite{Tch_fun}:
			\begin{equation}
				\begin{split}
				\ln Z^{\mathrm{G C}}\left(T, V,\left\{\mu_{i}\right\}\right) =
				\sum_{\text {species\ i}} \frac{V g_{i}}{2 \pi^{2}} \int_{0}^{\infty} \pm p^{2} d p \\
				\times \ln \left[1 \pm \lambda_{i} \exp \left(-\beta \epsilon_{i}\right)\right],
				\end{split}
			\end{equation}
			where $g_{i}$ and $\mu_{i}$ are the degeneracy and chemical potential of the hadron species $i$, respectively, and $\beta = 1/T$ and
			$E_{i} = \sqrt{p^{2} + m_{i}^{2}}$ with $m_{i}$ being the mass of the particle. The upper sign corresponds to the fermions and the lower sign indicates the bosons, with fugacity $\lambda_{i}(T, \vec{\mu}) = \exp\left(\mu_{i}/{T}\right)$.
			The chemical potential for the particle species $i$ in this case is given by the following:
			\begin{equation}
				\mu_{i} = B_{i} \mu_{B} + Q_{i} \mu_{Q} + S_{i} \mu_{S},
			\end{equation}
			where $B_{i}$, $S_{i}$, and $Q_{i}$ are the baryon number, strangeness, and charge number, respectively, of the hadron species $i$,
			and $\mu_{B}$, $\mu_{Q}$, and $\mu_{S}$ are the respective chemical potentials. The particle multiplicities are given through the following:
			\begin{equation}
				\begin{aligned}
				N_{i}^{\mathrm{G C}} & = T \frac{\partial \ln Z^{\mathrm{G C}}}{\partial \mu_{i}} \\
				& = \frac{g_{i} V}{2 \pi^{2}} \sum_{k=1}^{\infty}(\mp 1)^{k+1} \frac{m_{i}^{2} T}{k} K_{2}\left(\frac{k m_{i}}{T}\right)
				e^{\beta k \mu_{i}},
				\end{aligned}
			\end{equation}
where $K_{2}$ is the Bessel function of the second order, $V = 4 / 3 \pi R_{i}^{3} $ is the hadron gas volume, and $R_{i}$ is the fireball radius.
			In the model, the resonances and their decay into lighter particles are important to the particle multiplicities:
			\begin{equation}
				\left\langle N_{i}\right\rangle(T, \left\{\mu_{i}\right\}) = \left\langle N_{i}\right\rangle^{t h}(T, \vec{\mu})+
				\sum_{j} \Gamma_{j \rightarrow i}\left\langle N_{j}\right\rangle^{t h, R}(T, \vec{\mu}),
			\end{equation}
			where the first term describes the thermal average particle multiplicity of species $i$,
			and the second term describes the overall resonance contributions to the particle multiplicity of the same species.

\begin{figure}[htb]
				\includegraphics[angle=0,scale=0.44]{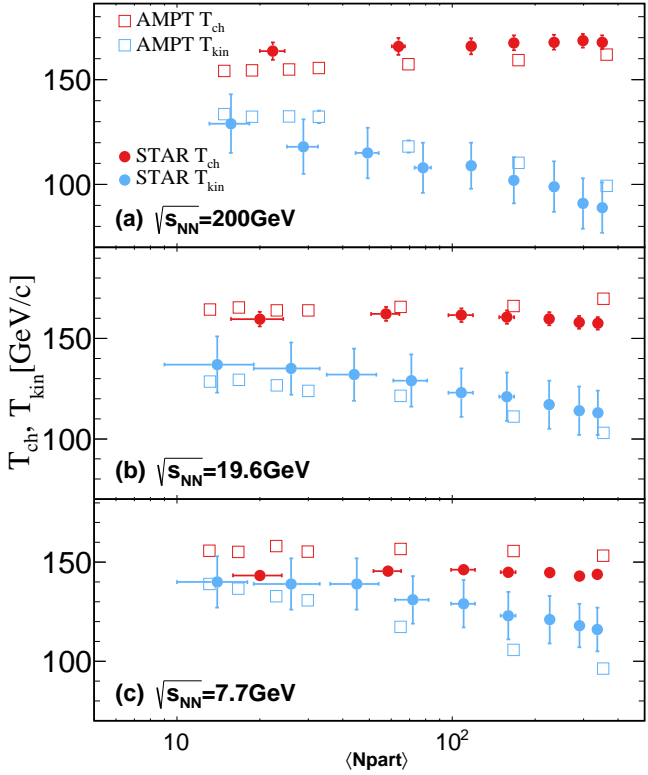}
				\caption{Chemical freeze-out parameters $T_{ch}$ and kinetic freeze-out parameters $T_{kin}$ versus $\left\langle \mathrm{N_{part}}\right\rangle$ in GCE from the fits to the particle yields with the $\sqrt{s_{NN}}$ = 200, 20, and 7.7 GeV AMPT simulations shown in (a), (b), and (c), respectively.
				The STAR data for $ \mathrm{Au} + \mathrm{Au}$ collisions at $\sqrt{s_{NN}} $ = 200, 19.6, and 7.7 GeV are taken from Ref.~\cite{09STAR_bulk,17STAR_bulk}.}
				\label{Tch_Tkin_new}
			\end{figure}

\begin{figure*}[htb]
				\includegraphics[angle=0,scale=0.9]{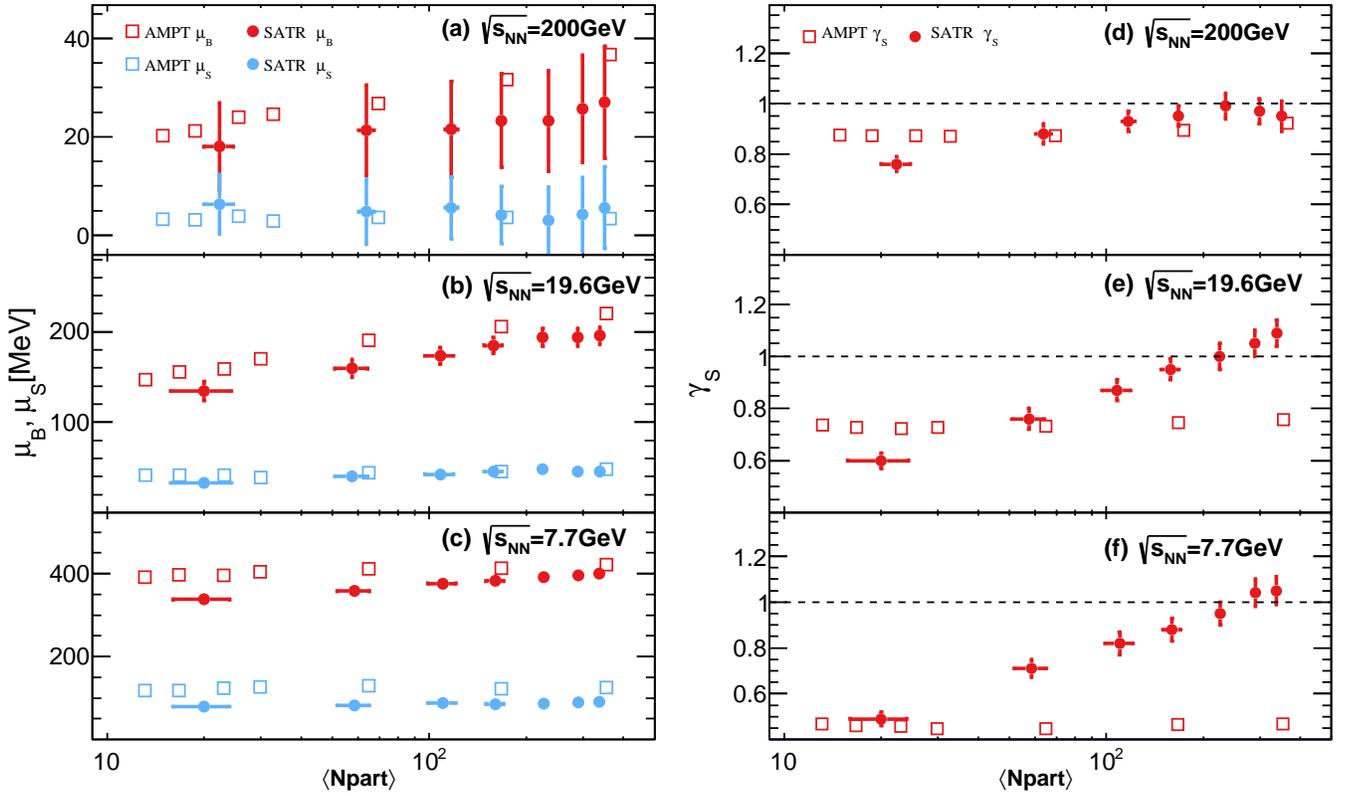}
				\caption{Chemical freeze-out parameters $\mu_{B}$, $\mu_{S}$, and $\gamma_{S}$ versus	 $\left\langle \mathrm{N_{part}}\right\rangle$ in GCE from the fits to the particle yields with $\sqrt{s_{NN}}$ = 200, 20, and 7.7 GeV during the AMPT simulation.
				The STAR data for $\mathrm{Au} + \mathrm{Au}$ collisions at $\sqrt{s_{NN}}$ = 200, 19.6, and 7.7 GeV are taken from Ref.~\cite{17STAR_bulk}.}
				\label{mub_mus_same_new}
			\end{figure*}

\begin{figure}[htb]
				\includegraphics[angle=0,scale=0.44]{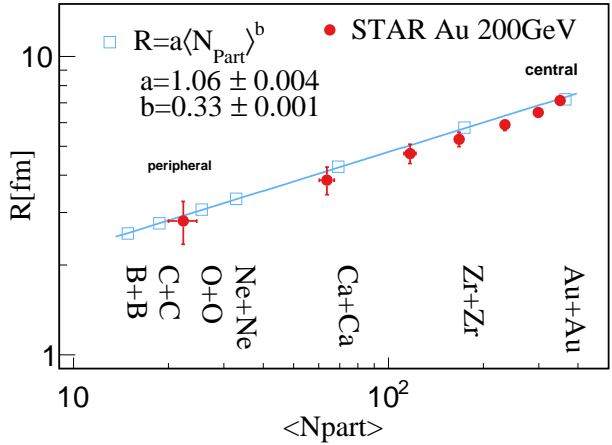}
				\caption{The radius parameter $R$ versus the average multiplicity $\left\langle \mathrm{N_{part}}\right\rangle$. The best fit with the $\sqrt{s_{NN}}$ = 200 GeV AMPT simulation results gives $R = a \left(\left\langle \mathrm{N_{part}}\right\rangle\right)^{b} $, where $a$ = 1.0 fm,
				and $b \approx 0.33$.
				A star point represents the STAR Collaboration data for $\mathrm{Au} + \mathrm{Au}$ at $\sqrt{s_{NN}}$= 200 GeV ~\cite{17STAR_bulk}.}
						\label{V_Npart}
			\end{figure}

			The chemical freeze-out temperature $T_{ch}$, baryon chemical potential $\mu_{B}$,
			strange chemical potential $\mu_{S}$, strangeness suppression factor $\gamma_{S}$
			(an allowance for a possibly incomplete strangeness equilibration is made by multiplying this factor for each particle species~\cite{gammas_2, gammas_1}), and canonical radius parameter $R$ are shown in Fig.~\ref{Tch_Tkin_new}, \ref{mub_mus_same_new}, and \ref{V_Npart}, respectively, for different collision systems and at different energies.
			\par
		
			In these figures, the extracted chemical freeze-out parameters of $\mathrm{{}^{197}Au} + \mathrm{{}^{197}Au}$ collisions at $\sqrt{s_{NN}} = 200$ GeV can match the experimental measurements conducted by the RHIC-STAR Collaboration~\cite{17STAR_bulk}.
			
			\par
			As Fig.~\ref{Tch_Tkin_new} (a) shows,
			at $\sqrt{s_{NN}}$ = 200 GeV,
			$T_{ch}$ as a function of $N_{part}$ for different collision systems presents a slight upward trend, whereas the kinetic freeze-out temperature $T_{kin}$ takes an inverse trend, which implies that a larger system reaches the chemical freeze-out stage with a higher temperature but with a significant expansion for a kinetic freeze-out.
			Meanwhile, we can observe that our fitting result in a $ \mathrm{{}^{197}Au} + \mathrm{{}^{197}Au}$ collision system is in accordance with a monotonically increasing function of $N_{part}$, extracted from $0\%-5\%$ central $ \mathrm{Au} + \mathrm{Au}$ collisions during the RHIC experiments~\cite{17STAR_bulk}.
			We also checked the results at $\sqrt{s_{NN}}$ = 20 and 7.7 GeV, as plotted in Fig.~\ref{Tch_Tkin_new} (b) and (c).
			Compared with those three energies, $T_{ch}$ has a weak dependence on $\left\langle \mathrm{N_{part}}\right\rangle$, whereas $T_{kin}$ clearly decreases as $\left\langle \mathrm{N_{part}}\right\rangle$ increases.
			As we can see, at a given system and different energy, $T_{ch}$ is approximately 160 MeV~\cite{17STAR_bulk}, which falls within a reasonable range of values given by the lattice QCD calculation.
			 At a small $\left\langle \mathrm{N_{part}}\right\rangle$, $T_{kin}$ is close to $T_{ch}$, but does not exceed the latter, which is a reasonable result based on a hydrodynamic assumption.

			\par
			Concerning the freeze-out potential, as shown in Fig.\ref{mub_mus_same_new}, the baryon chemical freeze-out potential $\mu_{B}$ increases with $N_{part}$, and the strangeness chemical freeze-out potential $\mu_{S}$ maintains a flat pattern with $N_{part}$, 	
			whereas the strangeness suppression factor $\gamma_{S}$ weakly increases with the collision system size and reaches close to unity for Au + Au collisions.
	For more details, Fig.~\ref{mub_mus_same_new} (a) demonstrates that $\mu_{B}$ increases from a small system ($ \mathrm{{}^{10}B} + \mathrm{{}^{10}B}$) to a larger system ($ \mathrm{{}^{197}Au} + \mathrm{{}^{197}Au}$),
			the behavior of which is qualitatively consistent with the fact that it increases from peripheral to central collisions at $\sqrt{s_{NN}}$ = 200 GeV during the STAR experiment~\cite{17STAR_bulk}.
			Our calculation plotted in Fig.~\ref{mub_mus_same_new} (b) shows that $\mu_{S}$ is almost constant from a small system to a larger system following the same behavior
			from peripheral to central collisions during the STAR experiment at $\sqrt{s_{NN}}$ = 200 GeV.  The strangeness suppression factor $\gamma_{S}$ is always lower than unity, which is also the case for those extracted from the STAR experiments at 200 GeV~\cite{17STAR_bulk}.
		
At $\sqrt{s_{NN}}$ = 20 and 7.7 GeV, as Fig.~\ref{mub_mus_same_new}(b) and (c) show, the results of $\mu_{B}$ and $\mu_{S}$ are also in line with the STAR results to a certain extent.
			However, the AMPT results in Fig.~\ref{mub_mus_same_new} (e) and (f) show that $\gamma_{S}$ remains constant, whereas the STAR $\gamma_{S}$ results show an increasing trend with $\left\langle \mathrm{N_{part}}\right\rangle$. This suggests that the mechanism of the strangeness production should be improved with this model.
		
			\par
			The radius parameter $R$ is related to the volume of the fireball at a chemical freeze-out and is obtained for the fitting yields. As shown in Fig.~\ref{V_Npart}, for $\sqrt{s_{NN}}$ = 200 GeV, $R$ indicates a strong $\left\langle \mathrm{N_{part}}\right\rangle$ dependence with a pattern of $R = a \left(\left\langle \mathrm{N_{part}}\right\rangle\right)^{b}$. The exponent of $b$ approximately equals $1/3$, and the coefficient $a$ is close to $1.0$ fm. Furthermore, we can see that the radius parameter $R$
			extracted from the STAR data from the peripheral to the central $ \mathrm{Au} + \mathrm{Au}$ collisions at $\sqrt{s_{NN}} = 200$ GeV, as represented by the circles,
			is close to our fitted line, showing a strong $\left\langle \mathrm{N_{part}}\right\rangle$ dependence.

			In contrast with $\sqrt{s_{NN}}$ = 200 GeV, the results at 20 and 7.7 GeV indicate a similar conclusion, but with slightly different parameter $a$ values of 0.8 and 0.7 fm, respectively, whereas parameter $b$ remains constant at approximately 1/3, which indicates that the number of participants is always proportional to the fireball volume.		
						
			\par
	\subsection{Nuclear modification factor with respect to $^{10}$B + $^{10}$B system}
	
			\begin{figure}[htb]
			\vspace{0.5cm}
				\includegraphics[angle=0,scale=0.44]{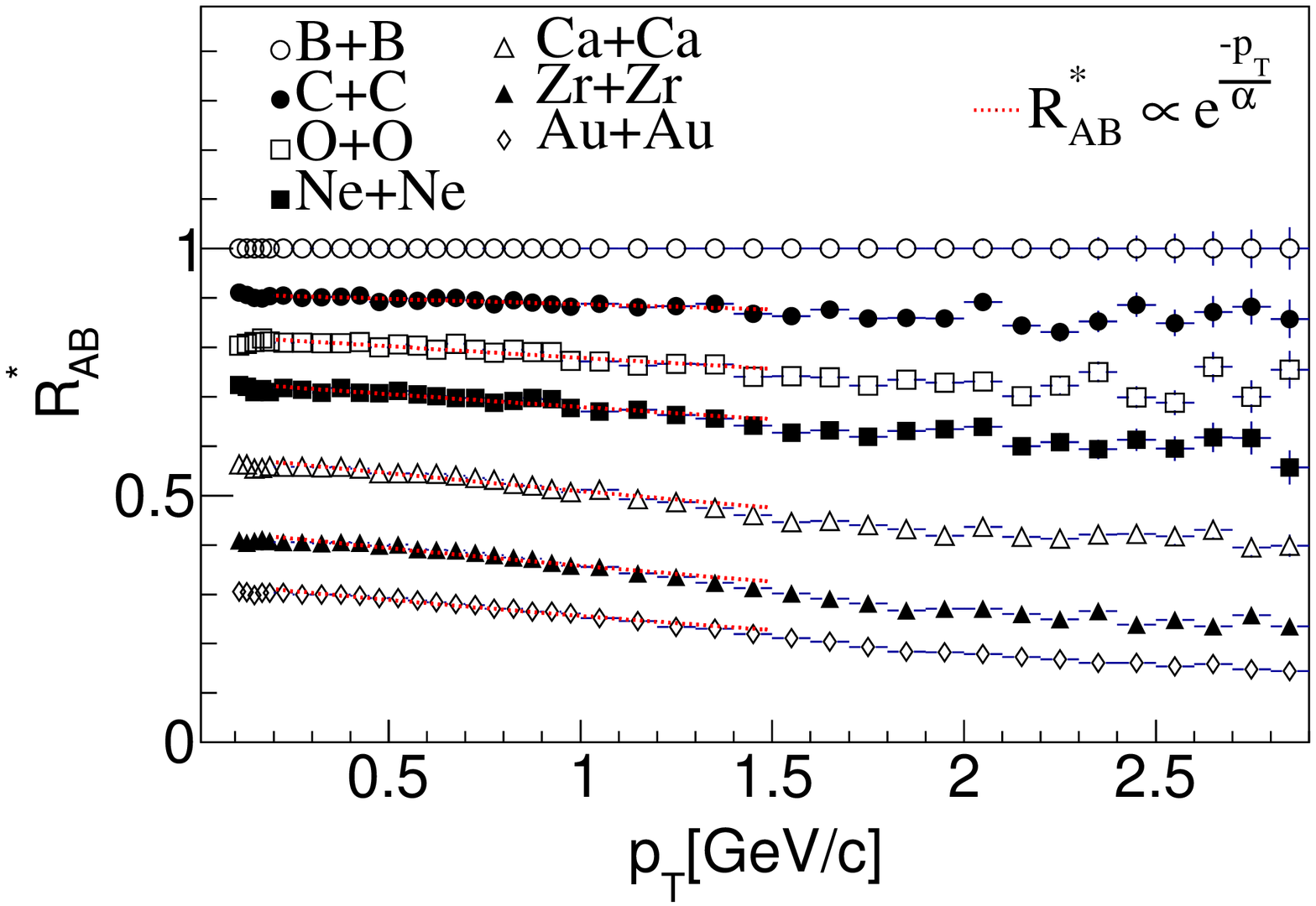}
				\caption{$R^*_{AB}$ as a function of $p_{T}$ for $ \mathrm{{}^{12}C} + \mathrm{{}^{12}C}$,
			$ \mathrm{{}^{16}O} + \mathrm{{}^{16}O}$, $ \mathrm{{}^{20}Ne} + \mathrm{{}^{20}Ne}$, $ \mathrm{{}^{40}Ca} + \mathrm{{}^{40}Ca}$,
			$ \mathrm{{}^{96}Zr} + \mathrm{{}^{96}Zr}$, and $ \mathrm{{}^{197}Au} + \mathrm{{}^{197}Au}$ collisions at $\sqrt{s_{NN}} = 200$ GeV.}
				\label{Raa_ch_new}
			\end{figure}
			
			\begin{table}[]
			\scriptsize
			\centering
			\caption{$ R^*_{\text{AB}}$ fitting parameter for different collision systems through $R_{\text{AB}} \propto \exp(\frac{-p_{\perp}}{\alpha})$ fits within a $p_{T}$ range $0.2-1.5$ GeV/c}
			\label{Raa_fit}
			\centering 
				\begin{tabular}{ccc}	
		 			System & $1/\alpha$  \\
					\hline
					\midrule
    					$\mathrm{\leftidx{^{12}}C} + \mathrm{\leftidx{^{12}}C}$		&$\approx0.024$      \\
    					$\mathrm{\leftidx{^{16}}O}+\mathrm{\leftidx{^{16}}O}$		&$\approx0.058$    \\
    					$\mathrm{\leftidx{^{20}}Ne}+\mathrm{\leftidx{^{20}}Ne}$		&$\approx0.075$  \\
    					$\mathrm{\leftidx{^{40}}Ca}+\mathrm{\leftidx{^{40}}Ca}$		&$\approx0.14$     \\
    					$\mathrm{\leftidx{^{96}}Zr}+\mathrm{\leftidx{^{96}}Zr}$		&$\approx0.19$    \\
    					$\mathrm{\leftidx{^{197}}Au}+\mathrm{\leftidx{^{197}}Au}$ 	&$\approx0.24$    \\
					\hline
				\end{tabular}
			\end{table}

			To explore the particle production mechanism and the system size effect in different collision systems, we define a parameter $R^*_{\text{AB}}$, i.e.,
			 the ratio of the charge particle transverse momentum spectra of the systems of
			$ \mathrm{{}^{12}C} + \mathrm{{}^{12}C}$,
			$ \mathrm{{}^{16}O} + \mathrm{{}^{16}O}$,  $ \mathrm{{}^{20}Ne} + \mathrm{{}^{20}Ne}$, $ \mathrm{{}^{40}Ca} + \mathrm{{}^{40}Ca}$, 
			$ \mathrm{{}^{96}Zr} + \mathrm{{}^{96}Zr}$, and $ \mathrm{{}^{197}Au} + \mathrm{{}^{197}Au}$ collisions with respect to $ \mathrm{{}^{10}B} + \mathrm{{}^{10}B}$ collisions
		 scaled based on the number of nucleon-nucleon collisions $N_{coll}$, which is similar to the nuclear modification factor $R_{\text{CP}}$ or $R_{\text{AA}}$ of high-energy heavy-ion collision experiments~\cite{Raa_star}:
			\begin{equation}
				\begin{aligned}
				 R^*_{\text{AB}} = \frac{(d^{2}N_{AA}/dydp_{T})/N^{AA}_{coll}}{(d^{2}N_{BB}/dydp_{T})/N^{BB}_{coll}},
				\end{aligned}
			\end{equation}
			where $d^{2}N_{AA}/dydp_{T}$ are the transverse momentum spectra for $ \mathrm{{}^{12}C} + \mathrm{{}^{12}C}$,
			$ \mathrm{{}^{16}O} + \mathrm{{}^{16}O}$, $ \mathrm{{}^{20}Ne} + \mathrm{{}^{20}Ne}$, $ \mathrm{{}^{40}Ca} + \mathrm{{}^{40}Ca}$, 
			$ \mathrm{{}^{96}Zr} + \mathrm{{}^{96}Zr}$, and $ \mathrm{{}^{197}Au} + \mathrm{{}^{197}Au}$ collisions, and $d^{2}N_{BB}/dydp_{T}$ is the transverse momentum spectrum
			for $ \mathrm{{}^{10}B} + \mathrm{{}^{10}B}$ collisions.
			Figure~\ref{Raa_ch_new} shows $R^*_{\text{AB}}$ as a function of $p_{T}$ from the AMPT model with the string melting scenario in $ \mathrm{{}^{12}C} + \mathrm{{}^{12}C}$, $ \mathrm{{}^{16}O} + \mathrm{{}^{16}O}$, $ \mathrm{{}^{20}Ne} + \mathrm{{}^{20}Ne}$, $ \mathrm{{}^{40}Ca} + \mathrm{{}^{40}Ca}$, 
			$ \mathrm{{}^{96}Zr} + \mathrm{{}^{96}Zr}$, and $ \mathrm{{}^{197}Au} + \mathrm{{}^{197}Au}$ collisions systems.
			 It is clear that $R^*_{\text{AB}}$ is strongly suppressed in the $ \mathrm{{}^{197}Au} + \mathrm{{}^{197}Au}$ collision system in
			comparison with the $ \mathrm{{}^{12}C} + \mathrm{{}^{12}C}$ collision system.
			From $ \mathrm{{}^{12}C} + \mathrm{{}^{12}C}$ to $ \mathrm{{}^{197}Au} + \mathrm{{}^{197}Au}$ collision systems,  $R^*_{\text{AB}}$
			decreases with an increase in the transverse momentum, and the yield of the charge particles is suppressed at a high $p_{T}$
			with respect to the results from $ \mathrm{{}^{10}B} + \mathrm{{}^{10}B}$ collisions, which depict a stronger particle interaction in a larger sized system.

\par
			To describe this suppression phenomenon, a simple function was employed to fit $R^*_{\text{AB}}$ in different systems. As is known, the transverse momentum spectra
			can be roughly described through a $p_{T}$-exponential function~\cite{09STAR_bulk}, $\frac{dN}{p_{\perp}dp_{\perp}}\propto \exp(-p_{\perp}/T_{p_{\perp}})$, where $p_{\perp}$ is the transverse momentum
			and $T_{p_{\perp}}$ is the inverse slope of the spectra. Based on the definition of $R^*_{\text{AB}}$, $R^*_{\text{AB}}$ can be written
			as $R^*_{\text{AB}} \propto \exp(\frac{-p_{\perp}}{\alpha})$.
			Here, the parameter $\alpha = (T^{BB}_{\perp}T^{AA}_{\perp})/(T^{BB}_{\perp} - T^{AA}_{\perp})$ is related to the effective temperature $T_{p_{\perp}}$ in large and small systems,
			where $T^{AA}_{\perp}$ and $T^{BB}_{\perp}$ are the apparent temperatures in A + A collisions and $\mathrm{\leftidx{^{10}}B} + \mathrm{\leftidx{^{10}}B}$ collisions, respectively.
			Table~\ref{Raa_fit} shows the parameter $1/\alpha$ of this fitting, which has a monotonically increasing trend with the collision system size, indicating that the suppression
			is more significant in a large system than in a small system.

\section{summary}
\label{sec:summary}	
        	The particle yields and their $p_{T}$ spectra of $\pi$, $k$, and $p$ in $ \mathrm{{}^{10}B} + \mathrm{{}^{10}B}$, $ \mathrm{{}^{12}C} + \mathrm{{}^{12}C}$,
	$ \mathrm{{}^{16}O} + \mathrm{{}^{16}O}$, $ \mathrm{{}^{20}Ne} + \mathrm{{}^{20}Ne}$, $ \mathrm{{}^{40}Ca} + \mathrm{{}^{40}Ca}$, $ \mathrm{{}^{96}Zr} + \mathrm{{}^{96}Zr}$, and
	$ \mathrm{{}^{197}Au} + \mathrm{{}^{197}Au}$ collision systems at $\sqrt{s_{NN}} = 200$ GeV have been investigated using the AMPT model. The chemical freeze-out and kinetic freeze-out properties were studied based on the thermal equilibrium model and blast-wave model, respectively.
		For  $\mathrm{Au} + \mathrm{Au}$ collisions at $\sqrt{s_{NN}} = 200$ GeV, the extracted chemical and kinetic freeze-out parameters agree with the experimental measurements from the RHIC-STAR Collaboration~\cite{09STAR_bulk,17STAR_bulk}.	
        It was found that the kinetic freeze-out parameters $T_{kin}$ decrease with an increase in $\beta_{T}$, which agrees with the early finding of the STAR or ALICE data results.

These results also show little energy dependence. As the energy $\sqrt{s_{NN}}$ decreases, the discrepancies between the AMPT model $p_T$ spectra and the experimental data finally lead to a kinetic freeze-out temperature $T_{kin}$ lower than in the STAR data given at a similar radial flow velocity.

	For the chemical freeze-out parameters, the baryon chemical potential $\mu_{B}$ increases with $\left\langle \mathrm{N_{part}}\right\rangle$, the strangeness chemical potential $\mu_{S}$ remains constant with a deviation from $\left\langle \mathrm{N_{part}}\right\rangle$ of a few MeV in all collision systems, and $\gamma_{S}$ keeps the same value.
	The fireball radius $R$ has a strong $\left\langle \mathrm{N_{part}}\right\rangle$ dependence, and a rough formula of $a \left(\left\langle \mathrm{N_{part}}
        	\right\rangle\right)^{b}$ can fit it well. The values of this power-law formula coefficient $b$ are approximately equal to $1/3$, indicating that $\left\langle \mathrm{N_{part}}\right\rangle$ is directly proportional to the freeze-out volume of the fireball.
	
From an energy dependence perspective, it can be seen that, for a higher initial energy density originating from a higher energy collision, the system has larger chemical freeze-out temperatures $T_{ch}$, which are extracted from the thermal model fits. As the energy increases, the nuclear penetration becomes more significant. With this in mind, $\mu_{B}$ will gradually decrease along with $\mu_{S}$, which is consistent with the STAR data. However, the strangeness suppression factor $\gamma_{S}$ differs significantly from the SATR results, which indicates that the strangeness production mechanism in the model needs to be improved in the future.
	In addition, we can see that the nuclear modification factors with respect to the $ \mathrm{{}^{10}B} + \mathrm{{}^{10}B}$ collision system for different collision systems present a gradual
	suppression within the intermediate $p_{T}$ range from a small system to a larger system.
To summarize, our detailed study provides a reference for a system scan of the chemical and kinetic properties of hot and dense QCD matter created during heavy-ion collisions at wide RHIC beam scan energies.

	\begin{acknowledgements}
	
This work was supported in part by the National Natural Science Foundation of China under contract Nos. 11890714, 11875066, 11421505, and 11775288, and the National Key R\&D Program of China under Grant Nos. 2016YFE0100900 and 2018YFE0104600.

	\end{acknowledgements}
	\end{CJK*}	
		\bibliography{no}

\begin{thebibliography}{37}
\expandafter\ifx\csname natexlab\endcsname\relax\def\natexlab#1{#1}\fi
\expandafter\ifx\csname bibnamefont\endcsname\relax
  \def\bibnamefont#1{#1}\fi
\expandafter\ifx\csname bibfnamefont\endcsname\relax
  \def\bibfnamefont#1{#1}\fi
\expandafter\ifx\csname citenamefont\endcsname\relax
  \def\citenamefont#1{#1}\fi
\expandafter\ifx\csname url\endcsname\relax
  \def\url#1{\texttt{#1}}\fi
\expandafter\ifx\csname urlprefix\endcsname\relax\def\urlprefix{URL }\fi
\providecommand{\bibinfo}[2]{#2}
\providecommand{\eprint}[2][]{\url{#2}}

\bibitem[{\citenamefont{Adams~\textit{et al}.}(2005)}]{phase_e1}
\bibinfo{author}{\bibfnamefont{J.}~\bibnamefont{Adams~\textit{et al}.}},
  \bibinfo{journal}{Nuclear Physics A} \textbf{\bibinfo{volume}{757}},
  \bibinfo{pages}{102} (\bibinfo{year}{2005}),
  \urlprefix\url{http://dx.doi.org/10.1016/j.nuclphysa.2005.03.085}.

\bibitem[{\citenamefont{Back~\textit{et al}.}(2005)}]{phase_e2}
\bibinfo{author}{\bibfnamefont{B.~B.} \bibnamefont{Back~\textit{et al}.}},
  \bibinfo{journal}{Nuclear Physics A} \textbf{\bibinfo{volume}{757}},
  \bibinfo{pages}{28} (\bibinfo{year}{2005}),
  \urlprefix\url{http://dx.doi.org/10.1016/j.nuclphysa.2005.03.084}.

\bibitem[{\citenamefont{Adcox~\textit{et al}.}(2005)}]{phase_e3}
\bibinfo{author}{\bibfnamefont{K.}~\bibnamefont{Adcox~\textit{et al}.}},
  \bibinfo{journal}{Nuclear Physics A} \textbf{\bibinfo{volume}{757}},
  \bibinfo{pages}{184} (\bibinfo{year}{2005}),
  \urlprefix\url{http://dx.doi.org/10.1016/j.nuclphysa.2005.03.086}.

\bibitem[{\citenamefont{Arsene~\textit{et al}.}(2005)}]{phase_e4}
\bibinfo{author}{\bibfnamefont{I.}~\bibnamefont{Arsene~\textit{et al}.}},
  \bibinfo{journal}{Nuclear Physics A} \textbf{\bibinfo{volume}{757}},
  \bibinfo{pages}{1} (\bibinfo{year}{2005}),
  \urlprefix\url{http://dx.doi.org/10.1016/j.nuclphysa.2005.02.130}.

\bibitem[{\citenamefont{Braun-Munzinger
  et~al.}(2016)\citenamefont{Braun-Munzinger, Koch, Schafer, and
  Stachel}}]{PBMPhysRep}
\bibinfo{author}{\bibfnamefont{P.}~\bibnamefont{Braun-Munzinger}},
  \bibinfo{author}{\bibfnamefont{V.}~\bibnamefont{Koch}},
  \bibinfo{author}{\bibfnamefont{T.}~\bibnamefont{Schafer}}, \bibnamefont{and}
  \bibinfo{author}{\bibfnamefont{J.}~\bibnamefont{Stachel}},
  \bibinfo{journal}{Physics Reports} \textbf{\bibinfo{volume}{621}},
  \bibinfo{pages}{76 } (\bibinfo{year}{2016}),
  \urlprefix\url{https://doi.org/10.1016/j.physrep.2015.12.003}.

\bibitem[{\citenamefont{Chen et~al.}(2018)\citenamefont{Chen, Keane, Ma, Tang,
  and Xu}}]{ChenPhysRep}
\bibinfo{author}{\bibfnamefont{J.}~\bibnamefont{Chen}},
  \bibinfo{author}{\bibfnamefont{D.}~\bibnamefont{Keane}},
  \bibinfo{author}{\bibfnamefont{Y.-G.} \bibnamefont{Ma}},
  \bibinfo{author}{\bibfnamefont{A.}~\bibnamefont{Tang}}, \bibnamefont{and}
  \bibinfo{author}{\bibfnamefont{Z.}~\bibnamefont{Xu}},
  \bibinfo{journal}{Physics Reports} \textbf{\bibinfo{volume}{760}},
  \bibinfo{pages}{1} (\bibinfo{year}{2018}),
  \urlprefix\url{https://doi.org/10.1016/j.physrep.2018.07.002}.

\bibitem[{\citenamefont{Luo and Xu}(2017)}]{LuoNST}
\bibinfo{author}{\bibfnamefont{X.}~\bibnamefont{Luo}} \bibnamefont{and}
  \bibinfo{author}{\bibfnamefont{N.}~\bibnamefont{Xu}},
  \bibinfo{journal}{Nuclear Science and Techniques}
  \textbf{\bibinfo{volume}{28}}, \bibinfo{pages}{112} (\bibinfo{year}{2017}),
  \urlprefix\url{https://doi.org/10.1007/s41365-017-0257-0}.

\bibitem[{\citenamefont{Song et~al.}(2017)\citenamefont{Song, Zhou, and
  Gajdosova}}]{SongNST}
\bibinfo{author}{\bibfnamefont{H.}~\bibnamefont{Song}},
  \bibinfo{author}{\bibfnamefont{Y.}~\bibnamefont{Zhou}}, \bibnamefont{and}
  \bibinfo{author}{\bibfnamefont{K.}~\bibnamefont{Gajdosova}},
  \bibinfo{journal}{Nuclear Science and Techniques}
  \textbf{\bibinfo{volume}{28}}, \bibinfo{pages}{99} (\bibinfo{year}{2017}),
  \urlprefix\url{https://doi.org/10.1007/s41365-017-0245-4}.

\bibitem[{\citenamefont{Ma}(2019)}]{MaYG}
\bibinfo{author}{\bibfnamefont{Y.~G.} \bibnamefont{Ma}},
  \bibinfo{journal}{SCIENTIA SINICA Physica, Mechanica and Astronomica}
  \textbf{\bibinfo{volume}{49}}, \bibinfo{pages}{102001}
  (\bibinfo{year}{2019}), \urlprefix\url{https://doi.org/10.1360/SSPMA-
  2019-0236}.

\bibitem[{\citenamefont{Andronic et~al.}(2018)\citenamefont{Andronic,
  Braun-Munzinger, Redlich, and Stachel}}]{PBMNature}
\bibinfo{author}{\bibfnamefont{A.}~\bibnamefont{Andronic}},
  \bibinfo{author}{\bibfnamefont{P.}~\bibnamefont{Braun-Munzinger}},
  \bibinfo{author}{\bibfnamefont{K.}~\bibnamefont{Redlich}}, \bibnamefont{and}
  \bibinfo{author}{\bibfnamefont{J.}~\bibnamefont{Stachel}},
  \bibinfo{journal}{Nature} \textbf{\bibinfo{volume}{561}},
  \bibinfo{pages}{321} (\bibinfo{year}{2018}),
  \urlprefix\url{https://doi.org/10.1038/s41586-018-0491-6}.

\bibitem[{\citenamefont{Aoki et~al.}(2009)\citenamefont{Aoki, Borsnyi, Drr,
  Fodor, Katz, Krieg, and Szabo}}]{tc_1}
\bibinfo{author}{\bibfnamefont{Y.}~\bibnamefont{Aoki}},
  \bibinfo{author}{\bibfnamefont{S.}~\bibnamefont{Borsnyi}},
  \bibinfo{author}{\bibfnamefont{S.}~\bibnamefont{Drr}},
  \bibinfo{author}{\bibfnamefont{Z.}~\bibnamefont{Fodor}},
  \bibinfo{author}{\bibfnamefont{S.~D.} \bibnamefont{Katz}},
  \bibinfo{author}{\bibfnamefont{S.}~\bibnamefont{Krieg}}, \bibnamefont{and}
  \bibinfo{author}{\bibfnamefont{K.}~\bibnamefont{Szabo}},
  \bibinfo{journal}{Journal of High Energy Physics}
  \textbf{\bibinfo{volume}{2009}}, \bibinfo{pages}{088} (\bibinfo{year}{2009}),
  \urlprefix\url{http://dx.doi.org/10.1088/1126-6708/2009/06/088}.

\bibitem[{\citenamefont{Bazavov et~al.}(2012)\citenamefont{Bazavov,
  Bhattacharya, Cheng, DeTar, Ding, Gottlieb, Gupta, Hegde, Heller, Karsch
  et~al.}}]{tc_2}
\bibinfo{author}{\bibfnamefont{A.}~\bibnamefont{Bazavov}},
  \bibinfo{author}{\bibfnamefont{T.}~\bibnamefont{Bhattacharya}},
  \bibinfo{author}{\bibfnamefont{M.}~\bibnamefont{Cheng}},
  \bibinfo{author}{\bibfnamefont{C.}~\bibnamefont{DeTar}},
  \bibinfo{author}{\bibfnamefont{H.-T.} \bibnamefont{Ding}},
  \bibinfo{author}{\bibfnamefont{S.}~\bibnamefont{Gottlieb}},
  \bibinfo{author}{\bibfnamefont{R.}~\bibnamefont{Gupta}},
  \bibinfo{author}{\bibfnamefont{P.}~\bibnamefont{Hegde}},
  \bibinfo{author}{\bibfnamefont{U.~M.} \bibnamefont{Heller}},
  \bibinfo{author}{\bibfnamefont{F.}~\bibnamefont{Karsch}},
  \bibnamefont{et~al.}, \bibinfo{journal}{Physical Review D}
  \textbf{\bibinfo{volume}{85}} (\bibinfo{year}{2012}),
  \urlprefix\url{http://dx.doi.org/10.1103/PhysRevD.85.054503}.

\bibitem[{\citenamefont{Fodor and Katz}(2004)}]{QCD2}
\bibinfo{author}{\bibfnamefont{Z.}~\bibnamefont{Fodor}} \bibnamefont{and}
  \bibinfo{author}{\bibfnamefont{S.~D.} \bibnamefont{Katz}},
  \bibinfo{journal}{JHEP} \textbf{\bibinfo{volume}{04}}, \bibinfo{pages}{050}
  (\bibinfo{year}{2004}).

\bibitem[{\citenamefont{Abelev~\textit{et al}}(2009)}]{09STAR_bulk}
\bibinfo{author}{\bibfnamefont{B.~I.} \bibnamefont{Abelev~\textit{et al}}}
  (\bibinfo{collaboration}{STAR Collaboration}), \bibinfo{journal}{Phys. Rev.
  C} \textbf{\bibinfo{volume}{79}}, \bibinfo{pages}{034909}
  (\bibinfo{year}{2009}),
  \urlprefix\url{https://link.aps.org/doi/10.1103/PhysRevC.79.034909}.

\bibitem[{\citenamefont{Abelev~\textit{et al}.}(2013)}]{13ALICE_276}
\bibinfo{author}{\bibfnamefont{B.}~\bibnamefont{Abelev~\textit{et al}.}}
  (\bibinfo{collaboration}{ALICE Collaboration}), \bibinfo{journal}{Phys. Rev.
  C} \textbf{\bibinfo{volume}{88}}, \bibinfo{pages}{044910}
  (\bibinfo{year}{2013}),
  \urlprefix\url{https://link.aps.org/doi/10.1103/PhysRevC.88.044910}.

\bibitem[{\citenamefont{Lao et~al.}(2018{\natexlab{a}})\citenamefont{Lao, Liu,
  Li, Duan, and Lacey}}]{Tfree}
\bibinfo{author}{\bibfnamefont{H.~L.} \bibnamefont{Lao}},
  \bibinfo{author}{\bibfnamefont{F.~H.} \bibnamefont{Liu}},
  \bibinfo{author}{\bibfnamefont{B.~C.} \bibnamefont{Li}},
  \bibinfo{author}{\bibfnamefont{M.-Y.} \bibnamefont{Duan}}, \bibnamefont{and}
  \bibinfo{author}{\bibfnamefont{R.~A.} \bibnamefont{Lacey}},
  \bibinfo{journal}{Nuclear Science and Techniques}
  \textbf{\bibinfo{volume}{29}}, \bibinfo{pages}{164}
  (\bibinfo{year}{2018}{\natexlab{a}}),
  \urlprefix\url{https://doi.org/10.1007/s41365-018-0504-z}.

\bibitem[{\citenamefont{Lao et~al.}(2018{\natexlab{b}})\citenamefont{Lao, Liu,
  Li, and Duan}}]{Tfree2}
\bibinfo{author}{\bibfnamefont{H.-L.} \bibnamefont{Lao}},
  \bibinfo{author}{\bibfnamefont{F.-H.} \bibnamefont{Liu}},
  \bibinfo{author}{\bibfnamefont{B.-C.} \bibnamefont{Li}}, \bibnamefont{and}
  \bibinfo{author}{\bibfnamefont{M.-Y.} \bibnamefont{Duan}},
  \bibinfo{journal}{Nuclear Science and Techniques}
  \textbf{\bibinfo{volume}{29}}, \bibinfo{pages}{82}
  (\bibinfo{year}{2018}{\natexlab{b}}),
  \urlprefix\url{https://doi.org/10.1007/s41365-018-0425-x}.

\bibitem[{\citenamefont{Braun-Munzinger
  et~al.}(1996)\citenamefont{Braun-Munzinger, Stachel, Wessels, and
  Xu}}]{sps_thermal}
\bibinfo{author}{\bibfnamefont{P.}~\bibnamefont{Braun-Munzinger}},
  \bibinfo{author}{\bibfnamefont{J.}~\bibnamefont{Stachel}},
  \bibinfo{author}{\bibfnamefont{J.}~\bibnamefont{Wessels}}, \bibnamefont{and}
  \bibinfo{author}{\bibfnamefont{N.}~\bibnamefont{Xu}},
  \bibinfo{journal}{Physics Letters B} \textbf{\bibinfo{volume}{365}},
  \bibinfo{pages}{1} (\bibinfo{year}{1996}),
  \urlprefix\url{http://dx.doi.org/10.1016/0370-2693(95)01258-3}.

\bibitem[{\citenamefont{Becattini et~al.}(2006)\citenamefont{Becattini,
  Manninen, and Ga\ifmmode~\acute{z}\else \'{z}\fi{}dzicki}}]{otherssds}
\bibinfo{author}{\bibfnamefont{F.}~\bibnamefont{Becattini}},
  \bibinfo{author}{\bibfnamefont{J.}~\bibnamefont{Manninen}}, \bibnamefont{and}
  \bibinfo{author}{\bibfnamefont{M.}~\bibnamefont{Ga\ifmmode~\acute{z}\else
  \'{z}\fi{}dzicki}}, \bibinfo{journal}{Phys. Rev. C}
  \textbf{\bibinfo{volume}{73}}, \bibinfo{pages}{044905}
  (\bibinfo{year}{2006}),
  \urlprefix\url{https://doi.org/10.1103/PhysRevC.73.044905}.

\bibitem[{\citenamefont{Citron~\textit{et al}.}(2018)}]{sysscanrhic}
\bibinfo{author}{\bibfnamefont{Z.}~\bibnamefont{Citron~\textit{et al}.}},
  \emph{\bibinfo{title}{Future physics opportunities for high-density qcd at
  the lhc with heavy-ion and proton beams}} (\bibinfo{year}{2018}),
  \eprint{1812.06772}.

\bibitem[{\citenamefont{Lin et~al.}(2005)\citenamefont{Lin, Ko, Li, Zhang, and
  Pal}}]{AMPT_origin}
\bibinfo{author}{\bibfnamefont{Z.-W.} \bibnamefont{Lin}},
  \bibinfo{author}{\bibfnamefont{C.~M.} \bibnamefont{Ko}},
  \bibinfo{author}{\bibfnamefont{B.-A.} \bibnamefont{Li}},
  \bibinfo{author}{\bibfnamefont{B.}~\bibnamefont{Zhang}}, \bibnamefont{and}
  \bibinfo{author}{\bibfnamefont{S.}~\bibnamefont{Pal}},
  \bibinfo{journal}{Phys. Rev. C} \textbf{\bibinfo{volume}{72}},
  \bibinfo{pages}{064901} (\bibinfo{year}{2005}),
  \urlprefix\url{https://link.aps.org/doi/10.1103/PhysRevC.72.064901}.

\bibitem[{\citenamefont{Xu and Ko}(2011)}]{xujun}
\bibinfo{author}{\bibfnamefont{J.}~\bibnamefont{Xu}} \bibnamefont{and}
  \bibinfo{author}{\bibfnamefont{C.~M.} \bibnamefont{Ko}},
  \bibinfo{journal}{Phys. Rev. C} \textbf{\bibinfo{volume}{83}},
  \bibinfo{pages}{034904} (\bibinfo{year}{2011}),
  \urlprefix\url{https://link.aps.org/doi/10.1103/PhysRevC.83.034904}.

\bibitem[{\citenamefont{Pal and Bleicher}(2012)}]{suppressionhighpt}
\bibinfo{author}{\bibfnamefont{S.}~\bibnamefont{Pal}} \bibnamefont{and}
  \bibinfo{author}{\bibfnamefont{M.}~\bibnamefont{Bleicher}},
  \emph{\bibinfo{title}{Suppression of high transverse momenta hadrons in pb+pb
  collisions at lhc}} (\bibinfo{year}{2012}),
  \urlprefix\url{http://dx.doi.org/10.1016/j.physletb.2012.01.070}.

\bibitem[{\citenamefont{Ye et~al.}(2017)\citenamefont{Ye, Chen, Ma, Zhang, and
  Zhong}}]{Ye_2017}
\bibinfo{author}{\bibfnamefont{Y.~J.} \bibnamefont{Ye}},
  \bibinfo{author}{\bibfnamefont{J.~H.} \bibnamefont{Chen}},
  \bibinfo{author}{\bibfnamefont{Y.~G.} \bibnamefont{Ma}},
  \bibinfo{author}{\bibfnamefont{S.}~\bibnamefont{Zhang}}, \bibnamefont{and}
  \bibinfo{author}{\bibfnamefont{C.}~\bibnamefont{Zhong}},
  \bibinfo{journal}{Chinese Physics C} \textbf{\bibinfo{volume}{41}},
  \bibinfo{pages}{084101} (\bibinfo{year}{2017}),
  \urlprefix\url{http://dx.doi.org/10.1088/1674-1137/41/8/084101}.

\bibitem[{\citenamefont{Jin et~al.}(2018)\citenamefont{Jin, Chen, Ma, Zhang,
  Zhang, and Zhong}}]{JinXH}
\bibinfo{author}{\bibfnamefont{X.~H.} \bibnamefont{Jin}},
  \bibinfo{author}{\bibfnamefont{J.~H.} \bibnamefont{Chen}},
  \bibinfo{author}{\bibfnamefont{Y.~G.} \bibnamefont{Ma}},
  \bibinfo{author}{\bibfnamefont{S.}~\bibnamefont{Zhang}},
  \bibinfo{author}{\bibfnamefont{C.~J.} \bibnamefont{Zhang}}, \bibnamefont{and}
  \bibinfo{author}{\bibfnamefont{C.}~\bibnamefont{Zhong}},
  \bibinfo{journal}{Nuclear Science and Techniques}
  \textbf{\bibinfo{volume}{29}}, \bibinfo{pages}{54} (\bibinfo{year}{2018}),
  \urlprefix\url{https://doi.org/10.1007/s41365-018-0393-1}.

\bibitem[{\citenamefont{Wang et~al.}(2019)\citenamefont{Wang, Chen, Ma, and
  Zhang}}]{WangH}
\bibinfo{author}{\bibfnamefont{H.}~\bibnamefont{Wang}},
  \bibinfo{author}{\bibfnamefont{J.-H.} \bibnamefont{Chen}},
  \bibinfo{author}{\bibfnamefont{Y.-G.} \bibnamefont{Ma}}, \bibnamefont{and}
  \bibinfo{author}{\bibfnamefont{S.}~\bibnamefont{Zhang}},
  \bibinfo{journal}{Nuclear Science and Techniques}
  \textbf{\bibinfo{volume}{30}}, \bibinfo{pages}{185} (\bibinfo{year}{2019}),
  \urlprefix\url{https://doi.org/10.1007/s41365-019-0706-z}.

\bibitem[{\citenamefont{Wang and Gyulassy}(1991)}]{HIJING-1}
\bibinfo{author}{\bibfnamefont{X.-N.} \bibnamefont{Wang}} \bibnamefont{and}
  \bibinfo{author}{\bibfnamefont{M.}~\bibnamefont{Gyulassy}},
  \bibinfo{journal}{Phys. Rev. D} \textbf{\bibinfo{volume}{44}},
  \bibinfo{pages}{3501} (\bibinfo{year}{1991}),
  \urlprefix\url{https://doi.org/10.1103/PhysRevD.44.3501}.

\bibitem[{\citenamefont{Gyulassy and Wang}(1994)}]{HIJING-2}
\bibinfo{author}{\bibfnamefont{M.}~\bibnamefont{Gyulassy}} \bibnamefont{and}
  \bibinfo{author}{\bibfnamefont{X.-N.} \bibnamefont{Wang}},
  \bibinfo{journal}{Computer Physics Communications}
  \textbf{\bibinfo{volume}{83}}, \bibinfo{pages}{307 } (\bibinfo{year}{1994}),
  \urlprefix\url{https://doi.org/10.1016/0010-4655(94)90057-4}.

\bibitem[{\citenamefont{Zhang}(1998)}]{ZPCModel}
\bibinfo{author}{\bibfnamefont{B.}~\bibnamefont{Zhang}},
  \bibinfo{journal}{Computer Physics Communications}
  \textbf{\bibinfo{volume}{109}}, \bibinfo{pages}{193 } (\bibinfo{year}{1998}),
  \urlprefix\url{https://doi.org/10.1016/S0010-4655(98)00010-1}.

\bibitem[{\citenamefont{Li and Ko}(1995)}]{ARTModel}
\bibinfo{author}{\bibfnamefont{B.-A.} \bibnamefont{Li}} \bibnamefont{and}
  \bibinfo{author}{\bibfnamefont{C.~M.} \bibnamefont{Ko}},
  \bibinfo{journal}{Phys. Rev. C} \textbf{\bibinfo{volume}{52}},
  \bibinfo{pages}{2037} (\bibinfo{year}{1995}),
  \urlprefix\url{https://doi.org/10.1103/PhysRevC.52.2037}.

\bibitem[{\citenamefont{Adler~\textit{et al}.}(2004)}]{PhysRevC.69.034909}
\bibinfo{author}{\bibfnamefont{S.~S.} \bibnamefont{Adler~\textit{et al}.}}
  (\bibinfo{collaboration}{PHENIX Collaboration}), \bibinfo{journal}{Phys. Rev.
  C} \textbf{\bibinfo{volume}{69}}, \bibinfo{pages}{034909}
  (\bibinfo{year}{2004}),
  \urlprefix\url{https://link.aps.org/doi/10.1103/PhysRevC.69.034909}.

\bibitem[{\citenamefont{Schnedermann et~al.}(1993)\citenamefont{Schnedermann,
  Sollfrank, and Heinz}}]{1993_ssBW}
\bibinfo{author}{\bibfnamefont{E.}~\bibnamefont{Schnedermann}},
  \bibinfo{author}{\bibfnamefont{J.}~\bibnamefont{Sollfrank}},
  \bibnamefont{and} \bibinfo{author}{\bibfnamefont{U.}~\bibnamefont{Heinz}},
  \bibinfo{journal}{Phys. Rev. C} \textbf{\bibinfo{volume}{48}},
  \bibinfo{pages}{2462} (\bibinfo{year}{1993}),
  \urlprefix\url{https://link.aps.org/doi/10.1103/PhysRevC.48.2462}.

\bibitem[{\citenamefont{Adamczyk~\textit{et al}.}(2017)}]{17STAR_bulk}
\bibinfo{author}{\bibfnamefont{L.}~\bibnamefont{Adamczyk~\textit{et al}.}}
  (\bibinfo{collaboration}{STAR Collaboration}), \bibinfo{journal}{Phys. Rev.
  C} \textbf{\bibinfo{volume}{96}}, \bibinfo{pages}{044904}
  (\bibinfo{year}{2017}),
  \urlprefix\url{https://link.aps.org/doi/10.1103/PhysRevC.96.044904}.

\bibitem[{\citenamefont{Braun-Munzinger
  et~al.}(2004)\citenamefont{Braun-Munzinger, Redlich, and Stachel}}]{Tch_fun}
\bibinfo{author}{\bibfnamefont{P.}~\bibnamefont{Braun-Munzinger}},
  \bibinfo{author}{\bibfnamefont{K.}~\bibnamefont{Redlich}}, \bibnamefont{and}
  \bibinfo{author}{\bibfnamefont{J.}~\bibnamefont{Stachel}},
  \emph{\bibinfo{title}{PARTICLE PRODUCTION IN HEAVY ION COLLISIONS}}
  (\bibinfo{publisher}{"World Scientific Publishing"}, \bibinfo{year}{2004}),
  pp. \bibinfo{pages}{491--599},
  \urlprefix\url{https://www.worldscientific.com/doi/abs/10.1142/9789812795533_0008}.

\bibitem[{\citenamefont{Becattini et~al.}(1998)\citenamefont{Becattini,
  Ga{\'{z}}dzicki, and Sollfrank}}]{gammas_2}
\bibinfo{author}{\bibfnamefont{F.}~\bibnamefont{Becattini}},
  \bibinfo{author}{\bibfnamefont{M.}~\bibnamefont{Ga{\'{z}}dzicki}},
  \bibnamefont{and}
  \bibinfo{author}{\bibfnamefont{J.}~\bibnamefont{Sollfrank}},
  \bibinfo{journal}{European Physical Journal C} \textbf{\bibinfo{volume}{5}},
  \bibinfo{pages}{143} (\bibinfo{year}{1998}),
  \urlprefix\url{https://doi.org/10.1007/s100529800831}.

\bibitem[{\citenamefont{Tawfik et~al.}(2015)\citenamefont{Tawfik, El-Bakry,
  Habashy, Mohamed, and Abbas}}]{gammas_1}
\bibinfo{author}{\bibfnamefont{A.~N.} \bibnamefont{Tawfik}},
  \bibinfo{author}{\bibfnamefont{M.~Y.} \bibnamefont{El-Bakry}},
  \bibinfo{author}{\bibfnamefont{D.~M.} \bibnamefont{Habashy}},
  \bibinfo{author}{\bibfnamefont{M.~T.} \bibnamefont{Mohamed}},
  \bibnamefont{and} \bibinfo{author}{\bibfnamefont{E.}~\bibnamefont{Abbas}},
  \bibinfo{journal}{International Journal of Modern Physics E}
  \textbf{\bibinfo{volume}{24}}, \bibinfo{pages}{1550067}
  (\bibinfo{year}{2015}),
  \urlprefix\url{https://doi.org/10.1142/S0218301315500676}.

\bibitem[{\citenamefont{Abelev~\textit{et al}}(2007)}]{Raa_star}
\bibinfo{author}{\bibfnamefont{B.~I.} \bibnamefont{Abelev~\textit{et al}}},
  \bibinfo{journal}{Physics Letters, Section B: Nuclear, Elementary Particle
  and High-Energy Physics} \textbf{\bibinfo{volume}{655}}, \bibinfo{pages}{104}
  (\bibinfo{year}{2007}),
  \urlprefix\url{https://doi.org/10.1016/j.physletb.2007.06.035}.

\end{thebibliography}

\end{document}